\documentclass[aps,prd, 10pt,nofootinbib, twocolumn]{revtex4}%
\usepackage{amsfonts}
\usepackage{amsmath}
\usepackage{amssymb}
\usepackage{graphicx}%
\providecommand{\U}[1]{\protect\rule{.1in}{.1in}}
\begin{document}
\title[Short title for running header]{Spatially inhomogeneous and irrotational geometries admitting Intrinsic
Conformal Symmetries}
\author{Pantelis S. Apostolopoulos$^{1}$}
\email{papost@phys.uoa.gr}
\affiliation{$^{1}$Technological Educational Institute of Ionian Islands, Department of
Environmental Technology, Panagoula 29100, Island of Zakynthos, Greece}
\pacs{PACS number}

\begin{abstract}
\textquotedblleft Diagonal\textquotedblright\ spatially inhomogeneous (SI)
models are introduced under the assumption of the existence of (proper)
intrinsic symmetries and can be seen, in some sense, complementary to the
Szekeres models. The structure of this class of spacetimes can be regarded as
a generalization of the (twist-free) Locally Rotationally Symmetric (LRS)
geometries without any global isometry containing, however, these models as
special cases. We consider geometries where a six-dimensional algebra
$\mathcal{IC}$ of Intrinsic Conformal Vector Fields (ICVFs) exists acting on a
$2-$dimensional (pseudo)-Riemannian manifold. Its members $\mathbf{X}_{\alpha
}$, constituted of 3 Intrinsic Killing Vector Fields (IKVFs) and 3
\emph{proper} and \emph{gradient} ICVFs, as well as the specific form of the
gravitational field are given explicitly. An interesting consequence, in
contrast with the Szekeres models, is the immediate existence of
\emph{conserved quantities along null geodesics}. We check computationally
that the magnetic part $H_{ab}$ of the Weyl tensor vanishes whereas the shear
$\sigma_{ab}$ and the electric part $E_{ab}$ share a common eigenframe
irrespective of the fluid interpretation of the models. A side result is the
fact that the spacetimes are foliated by a set of \emph{conformally flat }
$3-$dimensional \emph{timelike} slices when the anisotropy of the
\emph{flux-free} fluid is described only in terms of the 3 \emph{principal
inhomogeneous \textquotedblleft pressures\textquotedblright} $p_{\alpha}$ or
equivalently when the Ricci tensor shares the same basis of eigenvectors with
$\sigma_{ab}$ and $E_{ab}$. The conformal flatness also indicates that a
\emph{10-dimensional algebra} of ICVFs $\mathbf{\Xi}$ acting on the
$3-$dimensional timelike slices is highly possible to exist enriching in that
way the set of conserved quantities admitted by the SI models found in the
present paper.

\end{abstract}
\maketitle

\section{Introduction}

The inspection of the Einstein's Field Equations (EFEs)
\begin{equation}
G_{\ b}^{a}\equiv R_{\ b}^{a}-\frac{1}{2}R\delta_{\ b}^{a}=T_{\ b}^{a}
\label{FieldEquations1}%
\end{equation}
reveals the rich and strong correlation between the geometry of spacetime and
the dynamics. The latter is primarily encoded to a
realistic\footnote{Realistic implies that the dynamical portions of $T_{ab}$
must be derived from a set of well established phenomenological laws and not
by hand.} Energy-Momentum (EM) tensor $T_{ab}$. However, even if we assume
that the spacetime does not contain any dynamical fields, then $g_{ab}(x^{c})$
becomes itself a dynamical variable showing the complexity that arises from
this duality. It is thus evident that any intention to simplify $g_{ab}%
(x^{c})$ with some kind of symmetry must take into account the fusion between
the gravitational field and the spacetime geometry.

On the other hand observable quantities necessitate the existence of a unit
timelike vector field $u^{a}$ representing an average velocity
\cite{vanElst:1996dr} and its kinematical quantities $\theta$ (volume
expansion scalar), $\sigma_{ab}$ (anisotropic expansion trace-free tensor),
$\omega_{ba}$ (congruence's twist tensor), $\dot{u}^{a}$ (non-geodesic
indication 1-form) describe the distortion of the integral curves of $u^{a}$
as measured in the rest space of a \emph{comoving} observer%

\begin{align}
\theta &  \equiv u_{a;b}h^{ab},\hspace{0.4cm}\sigma_{ab}\equiv u_{(c;d)}%
\left(  h_{a}^{c}h_{b}^{d}-\frac{1}{3}h^{cd}h_{ab}\right) \nonumber\\
& \nonumber\\
\dot{u}_{a}  &  \equiv u_{a;b}u^{b},\hspace{0.4cm}\omega_{ba}\equiv
u_{[c;d]}h_{a}^{c}h_{b}^{d}. \label{TimelikeKinematical}%
\end{align}
where $h_{ab}=g_{ab}+u_{a}u_{b}$ is the projection tensor normally to $u^{a}$.
In the generic case there are no a priori reasons to impose special features
on the timelike congruence and only the interplay of physics (plus
observations) and geometry with the inclusion of appropriate boundary data (at
spatial or null past/future infinity) should enforce the need of such characteristics.

The third constituent element in this \textquotedblleft
arena\textquotedblright\ is the presence of a matter fluid which is described
in terms of the geometry and the kinematics as
\begin{equation}
T_{\ b}^{a}=\rho u^{a}u_{b}+ph_{\ b}^{a}+q^{a}u_{b}+u^{a}q_{b}+\pi_{\ b}^{a}
\label{EnergyMomentumTensor1}%
\end{equation}
where $\rho$, $p$ are the energy density and the isotropic pressure
respectively, $q^{a}$ is the direction of the momentum flow and $\pi_{ab}$ is
the anisotropic and trace-free pressure tensor%

\begin{align}
\rho &  \equiv T_{ab}u^{a}u^{b}\hspace{0.4cm}p\equiv\frac{1}{3}T_{ab}%
h^{ab}\nonumber\\
& \nonumber\\
q_{a}  &  \equiv-h_{a}^{c}T_{cd}u^{d}\hspace{0.4cm}\pi_{ab}\equiv\left(
h_{a}^{c}h_{b}^{d}-\frac{1}{3}h^{cd}h_{ab}\right)  T_{cd}.
\label{IrreducParts}%
\end{align}
Each of the above dynamical components (must) has a phenomenologically sound
meaning \cite{Weinberg} that can be justified from observations in some
acceptable cosmological scale. It should be noticed that the choice of the
observer is not unique and can be chosen either comoving $u^{a}$ or
non-comoving $\tilde{u}^{a}$($\neq u^{a}$) in which case the interpretation
for each one should be completely different leading to the notion of
\emph{tilted} models \cite{King:1972td}.

Spatially Inhomogeneous (SI) models \cite{Krasinski} provide a significant
work field towards to our understanding of the structure formation and the
effect of local density and pressure fluctuations in the accelerated phase of
the Universe. It is clear that they represent not an alternative of the
linearized version of the perturbed Friedmann-Lemaitre-Robertson-Walker (FLRW)
models but \emph{exact perturbation solutions} within a homogeneous and
isotropic background. Although, up to date, a quite generic SI model without
special characteristics (in the sense that will become transparent in the next
sections) has not be found, the known exact SI solutions can be served,
however, as toy models to various directions \cite{Krasinski:2013era}.

Szekeres solution \cite{Szekeres:1974ct} was the first SI model without any
(global) isometry and, as such, is well fitted along the aforementioned
research lines. From a geometrical and kinematical point of view, it admits a
tetrad of unit vector fields $\{u^{a},x^{a},y^{a},z^{a}\}$ that are
hypersurface orthogonal and any pair $\{u^{a},x^{a}\}$, $\{u^{a},y^{a}\}$,
$\{u^{a},z^{a}\}$ is surface forming which implies that
\begin{align}
y^{k}L_{\mathbf{u}}x_{k}  &  =z^{k}L_{\mathbf{u}}x_{k}=0\nonumber\\
& \nonumber\\
x^{k}L_{\mathbf{u}}y_{k}  &  =z^{k}L_{\mathbf{u}}y_{k}=0\nonumber\\
& \nonumber\\
x^{k}L_{\mathbf{u}}z_{k}  &  =y^{k}L_{\mathbf{u}}z_{k}.
\label{SurfaceForming1}%
\end{align}
In addition the unit timelike vector field $u^{a}$ is geodesic, consistent
with a dust fluid content (\cite{Szafron:1977zz, Szafron:1977zza} provide a
generalization of the Szekeres spacetime with $p\neq0$) which results to the
Szekeres family of quasi-symmetric models \cite{Hellaby:2002nx,
Apostolopoulos:2016nno}\begin{widetext}
\begin{equation}
ds^{2}=-dt^{2}+S^{2}\left\{  \frac{\left[  \left(  \ln S/E\right)  ^{\prime
}\right]  ^{2}}{\epsilon+F}dr^{2}+\frac{dy^{2}+dz^{2}}{V^{2}\left\{
1+\frac{k}{4}\left[  \left(  y-Y\right)  ^{2}+\left(  z-Z\right)  ^{2}\right]
\right\}  ^{2}}\right\}  \label{SzekeresForm1}%
\end{equation}%
\begin{equation}
ds^{2}=-dt^{2}+S^{2}\left\{  \frac{\left[  \left(  \ln S/E\right)  ^{\prime
}\right]  ^{2}}{F}dr^{2}+\frac{4\left(  dy^{2}+dz^{2}\right)  }{\left[
\left(  y-Y\right)  ^{2}+\left(  z-Z\right)  ^{2}\right]  ^{2}}\right\}
\label{SzekeresForm2}%
\end{equation}
\end{widetext}where $k=\epsilon/V^{2}$ and $Y(r),$ $Z(r),$ $V(r),$ $F(r)$ are
arbitrary functions of the radial coordinate. An important property of these
models is the vanishing of the magnetic part of the Weyl tensor
\begin{equation}
\frac{1}{2}\eta_{ac}^{\quad ij}\ C_{ijbd}u^{c}u^{d}\equiv H_{ab}=0
\label{MagneticPartWeyl1}%
\end{equation}
which implies that gravitational radiation cannot propagate
\cite{Maartens:1996hb, vanElst:1996zs} within this class of models.
Essentially equation (\ref{MagneticPartWeyl1}) is true for the general
diagonal metric ($u^{a}=C^{-1}\delta_{t}^{a}$)
\begin{equation}
ds^{2}=g_{ab}dx^{a}dx^{b}=A^{2}dx^{2}+B^{2}dz^{2}-C^{2}dt^{2}+D^{2}dy^{2}
\label{GeneralDiagonalMetric1}%
\end{equation}
therefore it can be seen entirely as an \textquotedblleft
artifact\textquotedblright\ of the specific geometrical character of the
tetrad $\{u^{a},x^{a},y^{a},z^{a}\}$ irrespective of further dynamical
restrictions. Spacetimes that satisfy equation (\ref{MagneticPartWeyl1}) are
usually referred as purely \textquotedblleft electrical\textquotedblright\ and
a lot of work has been done regarding the dynamical structure and the
existence of perfect fluid models (see e.g. \cite{Wylleman-Van den Bergh} and
references cited therein) with vanishing $H_{ab}$. The analysis is focused
mainly to perfect fluids with a barotropic equation of state $p=p(\rho)$ or
rotational dust (geodesic) models.

The key feature of the family (\ref{SzekeresForm1}) or (\ref{SzekeresForm2})
is the \emph{conformal flatness} of the $3-$dimensional slices $t=$const.
\cite{Hellaby:2002nx} which, geometrically, could be the reminiscent of the
constant\ curvature of the 2-dimensional hypersurfaces $t,r=$const. and the
subsequent existence of a \emph{6-dimensional algebra} of Intrinsic Conformal
Vector Fields (ICVFs) $\mathbf{X}$ satisfying \cite{Apostolopoulos:2016nno}
\begin{equation}
p_{a}^{c}p_{b}^{d}\mathcal{L}_{\mathbf{X}}p_{cd}=2\phi(\mathbf{X})p_{ab}
\label{ICVFs3Dimensional1}%
\end{equation}
where $p_{ab}=h_{ab}-x_{a}x_{b}$ is the projection tensor normal to the pair
$\{u^{a},x^{a}\}$ and, given the structure of (\ref{SzekeresForm1}) or
(\ref{SzekeresForm2}), represents the induced metric of the 2-dimensional
manifold $\mathbf{u}\wedge\mathbf{x}=\mathbf{0}$.

The notion of intrinsic symmetries has been introduced in \cite{Collins1,
Collins-Szafron1, Collins-Szafron2, Collins-Szafron3} without, however, giving
the covariant form of them. In order to investigate the implications of the
existence of geometric symmetries in general relativity we must take into
account the holonomy group structure of the spacetime manifold together with
the associated local diffeomorphisms \cite{Hall-Book}. Furthermore, it is
necessary to reformulate the necessary and sufficient (integrability)
conditions, coming from the existence of the symmetry, in a covariant way and
study their consequences in the kinematics and dynamics of the corresponding
model. The fact that Szekeres models admit (proper) ICVFs acting on
2-dimensional (and possibly 3-dimensional) submanifolds shows that ICVFs could
be more relevant and impose much less restrictions than the full CVFs-models
which are very rare \cite{Krasinski}.

The purpose of the present paper is to extent the investigation of the
existence of ICVFs to spacetimes with metric (\ref{GeneralDiagonalMetric1})
thus providing a some kind of \emph{geometrical} classification with respect
to the intrinsic conformal algebra \emph{without assuming} any matter content
thus providing a much richer diversity of possible physically sound models
than those that have been reported so far \cite{Wylleman-Van den Bergh}. In
particular, in Section II we assume that a \emph{6-dimensional algebra} of
ICVFs exists, acting on the timelike distribution $\mathbf{x}\wedge
\mathbf{z}=\mathbf{0}$ which implies that the latter has constant curvature
and the resulting spacetimes can be referred to as quasi-symmetric. We give
the explicit form of the ICVFs and the associated spacetime metrics and show
computationally that the magnetic part $H_{ab}$ of the Weyl tensor vanishes
whereas the shear $\sigma_{ab}$ and the electric part $E_{ab}=C_{acbd}%
u^{c}u^{d}$ share a common eigenframe irrespective of the fluid interpretation
of the models. Furthermore non-tilted perfect fluids (where, in general, $p$
and $\rho$ do not satisfy a barotropic equation of state) cannot be excluded
at once since the $H-$divergence constraint is trivially satisfied. Two
interesting results then arise: in contrast with the Szekeres models, there
exist \emph{infinite} \emph{conserved quantities} along null geodesics.
Furthermore the hypersurfaces $x=$const. are \emph{conformally flat} when the
fluid is flux-free $q^{a}=0$ and its anisotropy is described only in terms of
the 3 \emph{principal inhomogeneous \textquotedblleft
pressures\textquotedblright} $p_{\alpha}$ or, equivalently, when the Einstein
tensor $G_{\ b}^{a}$ is \textquotedblleft diagonal\textquotedblright. One
should expect the existence of \emph{10-dimensional algebra} of ICVFs
$\mathbf{\Xi}$ of the $\mathbf{x}_{\perp}-$distribution that satisfy
\begin{equation}
\hat{h}_{a}^{c}\hat{h}_{b}^{d}\mathcal{L}_{\mathbf{\Xi}}\hat{h}_{cd}%
=2\phi(\mathbf{\Xi})\hat{h}_{ab}\label{ICVFs3Dimension}%
\end{equation}
where $\hat{h}_{ab}=g_{ab}-x_{a}x_{b}$ is regarded as the induced metric of
$\mathbf{x}_{\perp}$. In section III, for completeness, we also give the
6-dimensional algebra of ICVFs acting on the $\mathbf{x}\wedge\mathbf{u}%
=\mathbf{0}$ spacelike distribution when $\dot{u}^{a}\neq0=x_{;b}^{a}x^{b}$.
As expected, the $x-$slices are also conformally flat provided that
$T_{\ b}^{a}=\mathrm{diag}(\rho,p_{1},p_{2},p_{3})$. Section IV includes our
conclusions and further areas of research.

Throughout this paper, the following conventions have been used: the spacetime
manifold is endowed with a Lorentzian metric of signature ($-,+,+,+$),
spacetime indices are denoted by lower case Latin letters $a,b,...=0,1,2,3$,
spatial frame indices are denoted by lower case Greek letters $\alpha
,\beta,...=1,2,3$ and we have used geometrized units such that $8\pi G=1=c$.

\section{Spatially Inhomogeneous and Irrotational models of type II}

We consider a spacetime geometry where a unit timelike vector field $u^{a}$ is
twist-free $\omega_{ab}=0$ but \emph{non-geodesic} $\dot{u}^{a}\neq0$. We make
the assumption that there exist 3 independent spacelike unit vector fields
$\left\{  \mathbf{x},\mathbf{y},\mathbf{z}\right\}  $, normal to $u^{a}$, and
each of these has the property to be hypersurface orthogonal
\begin{equation}
x_{[a}x_{b;c]}=y_{[a}y_{b;c]}=z_{[a}z_{b;c]}=0.
\label{HypersurfaceOrthogonal1}%
\end{equation}
The unit spacelike vector field $x^{a}$ is taken to be geodesic i.e. $\left(
x_{a}\right)  ^{\ast}\equiv x_{a;b}x^{b}=0$ and the pairs $\{u^{a},x^{a}\}$,
$\{u^{a},y^{a}\}$, $\{u^{a},z^{a}\}$ are surface forming satisfying eq.
(\ref{SurfaceForming1}).

Under these conditions the most general metric adapted to the geodesic
coordinates of $x^{a}$ has the following form
\begin{equation}
ds^{2}=g_{ab}dx^{a}dx^{b}=dx^{2}+B^{2}dz^{2}-C^{2}dt^{2}+D^{2}dy^{2}
\label{MetricTimelikeSurfaceConstantCurvature1}%
\end{equation}
where the functions $B(t,x,y,z)$, $C(t,x,y,z)$ and $D(t,x,y,z)$ depend on all
four coordinates. It follows from
(\ref{MetricTimelikeSurfaceConstantCurvature1}) that the magnetic part of the
Weyl tensor w.r.t. $u^{a}$ vanishes $H_{ab}=0$ and, in general, the Petrov
type is I that is $E_{ab}=\mathrm{diag}\left(  0,E_{1},E_{2},E_{3}\right)  $.

Essentially, the induced metric of the distribution $\mathbf{x}\wedge
\mathbf{z}=\mathbf{0}$ is represented by the second order symmetric tensor
$p_{ab}\equiv g_{ab}-x_{a}x_{b}-z_{a}z_{b}$ where $p_{a}^{k}x_{k}=0=p_{a}%
^{k}z_{k}$. We assume that there exist a 6-dimensional algebra $\mathcal{IC}%
(\mathbf{X}_{A})$ ($A=1,...,6$) of ICVFs acting on 2d pseudo-Riemannian
manifold that obey
\begin{equation}
p_{a}^{c}p_{b}^{d}\mathcal{L}_{\mathbf{X}}g_{cd}=p_{a}^{c}p_{b}^{d}%
\mathcal{L}_{\mathbf{X}}p_{cd}\equiv\bar{\nabla}_{(b}X_{a)}=2\phi
(\mathbf{X})p_{ab} \label{DefinitionICVFs1}%
\end{equation}
where $\phi(\mathbf{X}_{A})$ are the conformal factors of the vectors
$\mathbf{X}_{A}$ that are lying and acting on the submanifold $\mathbf{x}%
\wedge\mathbf{z}=\mathbf{0}$ and $\bar{\nabla}_{a}$ represents a well defined
covariant derivative
\begin{equation}
\bar{\nabla}_{c}p_{ab}=p_{c}^{k}p_{a}^{i}p_{b}^{j}\nabla_{k}p_{ij}=0
\label{TorsionFreeDerivative1}%
\end{equation}
for any tensorial quantity
\[
\bar{\nabla}_{c}\Pi_{b}^{a}\equiv p_{c}^{k}p_{i}^{a}p_{b}^{k}\Pi_{j;k}^{i}.
\]
From the inspection of equations (\ref{DefinitionICVFs1}) it follows that
$C=D$ and the general solution shows that $\mathbf{X}_{1},\mathbf{X}%
_{2},\mathbf{X}_{3}$ are \emph{Intrinsic Killing Vector Fields} (IKVFs) and
$\mathbf{X}_{4},\mathbf{X}_{5},\mathbf{X}_{6}$ are \emph{proper} and
\emph{gradient} ICVFs i.e. their associated bivectors vanish identically
$\bar{\nabla}_{[b}X_{a]}=0$
\begin{equation}
\mathbf{X}_{1}=\mathbf{M}_{yt}=(y-Y)\partial_{t}+\left(  t-T\right)
\partial_{y} \label{FullSpaceKVF1}%
\end{equation}%
\begin{align}
\mathbf{X}_{2}  &  =\left\{  \frac{k}{4}\left[  \left(  y-Y\right)
^{2}+\left(  t-T\right)  ^{2}\right]  -1\right\}  \partial_{t}+\nonumber\\
& \nonumber\\
&  +\frac{k}{2}\left(  y-Y\right)  \left(  t-T\right)  \partial_{y}
\label{FullSpaceKVF2}%
\end{align}%
\begin{align}
\mathbf{X}_{3}  &  =\frac{k}{2}\left(  y-Y\right)  \left(  t-T\right)
\partial_{t}+\nonumber\\
& \nonumber\\
&  +\left\{  1+\frac{k}{4}\left[  \left(  y-Y\right)  ^{2}+\left(  t-T\right)
^{2}\right]  \right\}  \partial_{y} \label{FullSpaceKVF3}%
\end{align}%
\begin{equation}
\mathbf{X}_{4}=\mathbf{H}=\left(  t-T\right)  \partial_{t}+\left(  y-Y\right)
\partial_{y} \label{FullSpaceCVF1}%
\end{equation}%
\begin{align}
\mathbf{X}_{5}  &  =\left\{  \frac{k}{4}\left[  \left(  t-T\right)
^{2}+\left(  y-Y\right)  ^{2}\right]  +1\right\}  \partial_{t}+\nonumber\\
& \nonumber\\
&  +\frac{k}{2}\left(  t-T\right)  \left(  y-Y\right)  \partial_{y}
\label{FullSpaceCVF2}%
\end{align}%
\begin{align}
\mathbf{X}_{6}  &  =\frac{k}{2}\left(  t-T\right)  \left(  y-Y\right)
\partial_{t}+\nonumber\\
& \nonumber\\
&  +\left\{  \frac{k}{4}\left[  \left(  y-Y\right)  ^{2}+\left(  t-T\right)
^{2}\right]  -1\right\}  \partial_{y}. \label{FullSpaceCVF3}%
\end{align}
with associated conformal factors
\begin{equation}
\phi(\mathbf{X}_{1})=\phi(\mathbf{X}_{2})=\phi(\mathbf{X}_{3})=0
\label{ConfFactorsLoerentzConstant1}%
\end{equation}%
\begin{equation}
\phi(\mathbf{X}_{4})=\left\{  1-\frac{k}{4}\left[  (y-Y)^{2}-(t-T)^{2}\right]
\right\}  N \label{ConfFactorsLoerentzConstant2}%
\end{equation}%
\begin{equation}
\phi(\mathbf{X}_{5})=kN(t-T),\quad\phi(\mathbf{X}_{6})=kN\left(  y-Y\right)  .
\label{ConfFactorsLoerentzConstant3}%
\end{equation}
Consequently the 2d manifold $\mathbf{x}\wedge\mathbf{z}=\mathbf{0}$ has
(locally) \emph{constant curvature} and the metric
(\ref{MetricTimelikeSurfaceConstantCurvature1}) takes the form
\begin{equation}
ds^{2}=dx^{2}+B^{2}dz^{2}+\frac{S^{2}}{V^{2}}\frac{-dt^{2}+dy^{2}}{\left\{
1+\frac{\epsilon}{4V^{2}}\left[  \left(  y-Y\right)  ^{2}-\left(  t-T\right)
^{2}\right]  \right\}  ^{2}} \label{MetricTimelikeSurfaceConstantCurvature2}%
\end{equation}
where $S(x,z)$, $Y(z),$ $T(z),$ $V(z)$ are arbitrary functions of their
arguments and $\epsilon=\pm1$ ($\neq0$) corresponds to the constant curvature
of the hypersurfaces $x,z=$const.

Defining the function $E(t,y,z)$ according to ($k=\epsilon/V^{2}$)
\begin{equation}
E(t,y,z)=V\left\{  1+\frac{k}{4}\left[  \left(  y-Y\right)  ^{2}-\left(
t-T\right)  ^{2}\right]  \right\}  \label{FunctionE1}%
\end{equation}
then
\begin{equation}
N(t,y,z)=\frac{1}{E(t,y,z)} \label{FunctionN1}%
\end{equation}
and the metric becomes
\begin{equation}
ds^{2}=dx^{2}+B^{2}dz^{2}+\frac{S^{2}}{E^{2}}\left(  -dt^{2}+dy^{2}\right)  .
\label{MetricTimelikeSurfaceConstantCurvature3}%
\end{equation}
The case where the distribution $\mathbf{x}\wedge\mathbf{z}=\mathbf{0}$ has
\emph{zero curvature} is treated similarly. The ICVFs are
\begin{equation}
\mathbf{X}_{1}=\mathbf{M}_{yt}=(y-Y)\partial_{t}+\left(  t-T\right)
\partial_{y} \label{FullSpaceFlatKVF1}%
\end{equation}%
\begin{align}
\mathbf{X}_{2}  &  =\left[  \left(  y-Y\right)  ^{2}+\left(  t-T\right)
^{2}\right]  \partial_{t}+\nonumber\\
& \nonumber\\
&  +2\left(  y-Y\right)  \left(  t-T\right)  \partial_{y}
\label{FullSpaceFlatKVF2}%
\end{align}%
\begin{align}
\mathbf{X}_{3}  &  =2\left(  y-Y\right)  \left(  t-T\right)  \partial
_{t}+\nonumber\\
& \nonumber\\
&  +\left[  \left(  y-Y\right)  ^{2}+\left(  t-T\right)  ^{2}\right]
\partial_{y} \label{FullSpaceFlatKVF3}%
\end{align}%
\begin{equation}
\mathbf{X}_{4}=\mathbf{H}=\left(  t-T\right)  \partial_{t}+\left(  y-Y\right)
\partial_{y} \label{FullSpaceFlatCVF1}%
\end{equation}%
\begin{equation}
\mathbf{X}_{5}=\partial_{t},\quad\mathbf{X}_{6}=\partial_{y}
\label{FullSpaceFlatCVF23}%
\end{equation}
with conformal factors
\begin{equation}
\phi(\mathbf{X}_{1})=\phi(\mathbf{X}_{2})=\phi(\mathbf{X}_{3})=0
\label{ConfFactorsLoerentzFlat1}%
\end{equation}%
\begin{equation}
\phi(\mathbf{X}_{4})=-1 \label{ConfFactorsLoerentzFlat2}%
\end{equation}%
\begin{equation}
\phi(\mathbf{X}_{5})=\frac{2(t-T)}{\left(  y-Y\right)  ^{2}-\left(
t-T\right)  ^{2}} \label{ConfFactorsLoerentzFlat3}%
\end{equation}%
\begin{equation}
\phi(\mathbf{X}_{6})=\frac{2(y-Y)}{\left(  t-T\right)  ^{2}-\left(
y-Y\right)  ^{2}} \label{ConfFactorsLoerentzFlat4}%
\end{equation}
and the metric function $E(t,y,z)$ is given by
\begin{equation}
E(t,y,z)=\frac{1}{N(t,y,z)}=\frac{1}{4}\left[  \left(  y-Y\right)
^{2}-\left(  t-T\right)  ^{2}\right]  . \label{FunctionE2}%
\end{equation}
A potential application of the IC algebra $\mathcal{IC}(\mathbf{X}_{A})$ found
in the present section could be the existence of conserved currents and
quantities. For example consider a null geodesic vector field $l^{a}$
\emph{lying} in the 2d manifold $\mathbf{x}\wedge\mathbf{z}=\mathbf{0}$ and
the quantities $Q_{A}=l^{a}X_{(A)a}$. It is easy to see that $Q_{A}$ are
conserved \emph{along} the null geodesics since
\begin{equation}
\left[  Q_{(A)}\right]  _{;a}l^{a}=l_{\ ;a}^{b}X_{(A)b}l^{a}+l^{a}%
l^{b}X_{(A)b;a}=0. \label{ConservedQuantities1}%
\end{equation}
For the metric (\ref{MetricTimelikeSurfaceConstantCurvature3}) a null geodesic
vector field is $l^{a}=f\left(  u^{a}+y^{a}\right)  =fn^{a}$ where $f(x^{a})$
satisfies $\left(  f_{;k}n^{k}\right)  n^{a}=-fn_{\ ;k}^{a}n^{k}$ (we note
that $n^{a}=u^{a}+y^{a}$ is not geodesic for a generic form of
(\ref{MetricTimelikeSurfaceConstantCurvature3})).

In the search for fluid solutions we usually start by analyzing the structure
of the constraints of the EFEs (\ref{FieldEquations1}). The \textquotedblleft
temporal\textquotedblright\ constraints $G_{\ \alpha}^{0}=0$ for the metric
(\ref{MetricTimelikeSurfaceConstantCurvature3}) reduce to
\begin{equation}
SB_{,tx}-B_{,t}S_{,x}=0 \label{Constraint1}%
\end{equation}%
\begin{equation}
B_{,y}E_{t}+B_{,t}E_{,y}+EB_{,ty}=0 \label{Constraint2}%
\end{equation}%
\begin{equation}
BS\left(  EE_{,zt}-E_{,t}E_{,z}\right)  +EB_{,t}\left(  ES_{,z}-SE_{,z}%
\right)  =0 \label{Constraint3}%
\end{equation}
whereas the \textquotedblleft spatial\textquotedblright\ constraints
$G_{\ \beta}^{\alpha}=0$ have the form
\begin{equation}
B_{,y}S_{,x}-SB_{,yx}=0 \label{Constraint4}%
\end{equation}%
\begin{equation}
B\left(  ES_{,zx}-S_{,x}E_{,z}\right)  +B_{,x}\left(  SE_{,z}-ES_{,z}\right)
=0 \label{Constraint5}%
\end{equation}%
\begin{equation}
BS\left(  EE_{,zy}-E_{,y}E_{,z}\right)  +EB_{,y}\left(  ES_{,z}-SE_{,z}%
\right)  =0 \label{Constraint6}%
\end{equation}
where a \textquotedblleft$,$\textquotedblright\ denotes partial
differentiation w.r.t. the corresponding coordinate.

The \emph{general solution} of the above set of coupled differential equations
is
\begin{equation}
B=\frac{S\left[  \ln\left(  S/E\right)  \right]  _{,z}}{\sqrt{\epsilon+F(z)}}
\label{SolutionFuncBTimelikeSurfConstantCurvat1}%
\end{equation}
where $F(z)$ is an arbitrary function and $E(t,y,z)$ is given in
(\ref{FunctionE1}) or (\ref{FunctionE2}).

It should be emphasized that the existence of the $\mathcal{IC}(\mathbf{X}%
_{A})$ intrinsic conformal algebra is a \emph{direct consequence} of the
general solution (\ref{SolutionFuncBTimelikeSurfConstantCurvat1}),
(\ref{FunctionE1}) or (\ref{FunctionE2}) therefore in order to determine the
exact form of $\mathbf{X}_{A}$ we could simply apply the methodology of
\cite{Apostolopoulos:2016nno} avoiding in that way eqs.
(\ref{DefinitionICVFs1}). Furthermore we can verify that the Petrov type is D
i.e. the eigenvalues of the electric part of the Weyl tensor $E_{1}=E_{3}$ (in
contrast with the Szekeres models where $E_{2}=E_{3}$).

The EFEs (\ref{FieldEquations1}) then become
\begin{equation}
G_{\ b}^{a}=T_{\ b}^{a}=\mathrm{diag}(\rho,p_{1},p_{2},p_{3})
\label{FieldEquations2}%
\end{equation}
i.e. the Ricci tensor $R_{\ b}^{a}$ \emph{shares the same basis of
eigenvectors} with $\sigma_{ab}$ and $E_{ab}$.

The \emph{directional} and \emph{inhomogeneous} \textquotedblleft
pressures\textquotedblright\ $p_{\alpha}$ are not necessarily equal and the
fluid is, in general, anisotropic for the comoving observers $u^{a}%
=(E/S)\delta_{\ t}^{a}$. In order to show if a specific perfect fluid solution
exists (i.e. $p_{1}=p_{2}=p_{3}$) one must monitor the integrability
conditions i.e. the consistent evolution of the \emph{non-trivial}
constraints. We can prove, however, that the div$-H$ constraint is trivially
satisfied. We observe computationally that the three mutually orthogonal and
unit spacelike vector fields $\left\{  x^{a},y^{a},z^{a}\right\}  $ are\emph{
}eigenvectors\emph{ }of $E_{ab}=\mathrm{diag}\left(  0,E_{1},E_{2}%
,E_{1}\right)  $ and $\sigma_{ab}=\mathrm{diag}\left(  0,\sigma_{1},\sigma
_{2},\sigma_{3}\right)  $. Because $H_{ab}$ vanishes identically for the
metric (\ref{MetricTimelikeSurfaceConstantCurvature3}) the further requirement
$p_{1}=p_{2}=p_{3}$ gives $\pi_{ab}=0$ and the $H-$divergence equation
\cite{vanElst:1996dr}%

\[
\epsilon^{\alpha\beta\gamma}\sigma_{\beta\delta}E_{\ \gamma}^{\delta}=0
\]
implies that the shear $\sigma_{ab}$ and the electric part $E_{ab}$ tensors
commute i.e. they must share a common eigenframe as actually do.

An important consequence of the solution (\ref{FunctionE1}) or
(\ref{FunctionE2}) and (\ref{SolutionFuncBTimelikeSurfConstantCurvat1}) is
that the Cotton-York tensor \cite{Garcia:2003bw, ValienteKroon:2004gj}
\begin{equation}
C_{abc}=2(R_{a[b}-\frac{1}{4}Rg_{a[b})_{;c]} \label{CottonYork1}%
\end{equation}
vanishes i.e. the hypersurfaces $x=$const. are \emph{conformally flat}.
Therefore in complete analogy with the 2d case, one should expect the
existence of \emph{10-dimensional algebra} of ICVFs $\mathbf{\Xi}$ of the
$\mathbf{x}_{\perp}-$distribution that satisfy
\begin{equation}
\hat{h}_{a}^{c}\hat{h}_{b}^{d}\mathcal{L}_{\mathbf{\Xi}}\hat{h}_{cd}%
=2\phi(\mathbf{\Xi})\hat{h}_{ab} \label{3dCVFsProperty1}%
\end{equation}
where $\hat{h}_{ab}=g_{ab}-x_{a}x_{b}$ is regarded as the induced metric of
$\mathbf{x}_{\perp}$.

We note that relaxing the flux-free restrictions (equations (\ref{Constraint1}%
)-(\ref{Constraint3})), exact perfect fluid models could be exist for
non-comoving (tilted) observers $\tilde{u}^{a}$ similar to the case of
Spatially Homogeneous (SH) tilted perfect models (e.g. \cite{Hervik:2007bb,
Hervik:2007uh, Apostolopoulos:2004fd}) which necessitates the presence of
\emph{non-zero vorticity} \cite{Jaffe:2005pw}. This could be also possible for
the Szekeres geometries i.e. when $\tilde{u}^{a}$ are \emph{comoving with the
(perfect) fluid} in which case the $u^{a}-$observers will interpret it as
imperfect\footnote{In \cite{Herrera:2012sc} the \textquotedblleft
environment\textquotedblright\ is completely different since the comoving
interpretation remains that of a perfect fluid (i.e. the exact Szekeres model)
and the tilted observers are derived from a Lorentz boost of $u^{a}$.}. Again,
if such a solution exists, it must be proved that evolves consistently along
$u^{a}$ that \textquotedblleft see\textquotedblright\ an anisotropic and
non-zero flux matter fluid.

\section{Spatially Inhomogeneous and Irrotational models of type III}

We are interested to the case where the induced metric of the distribution
$\mathbf{x}\wedge\mathbf{u}=\mathbf{0}$, represented by the second order
symmetric tensor $p_{ab}\equiv g_{ab}-x_{a}x_{b}+u_{a}u_{b}$ where $p_{a}%
^{k}x_{k}=0=p_{a}^{k}u_{k}$, admits the 6-dimensional algebra $\mathcal{IC}%
(\mathbf{X}_{A})$ ($A=1,...,6$) of ICVFs
\begin{equation}
p_{a}^{c}p_{b}^{d}\mathcal{L}_{\mathbf{X}}p_{cd}=2\phi(\mathbf{X})p_{ab}
\label{DefinitionICVFs2}%
\end{equation}%
\begin{equation}
\mathbf{X}_{1}=\mathbf{M}_{yz} \label{TypeIIIFullSpaceKVF1}%
\end{equation}%
\begin{align}
\mathbf{X}_{2}  &  =\left\{  1+\frac{k}{4}\left[  \left(  y-Y\right)
^{2}-\left(  z-Z\right)  ^{2}\right]  \right\}  \partial_{y}+\nonumber\\
& \nonumber\\
&  +\frac{k}{2}\left(  y-Y\right)  \left(  z-Z\right)  \partial_{z}
\label{TypeIIIFullSpaceKVF2}%
\end{align}%
\begin{align}
\mathbf{X}_{3}  &  =\frac{k}{2}\left(  y-Y\right)  \left(  z-Z\right)
\partial_{y}+\nonumber\\
& \nonumber\\
&  +\left\{  1+\frac{k}{4}\left[  \left(  z-Z\right)  ^{2}-\left(  y-Y\right)
^{2}\right]  \right\}  \partial_{z} \label{TypeIIIFullSpaceKVF3}%
\end{align}%
\begin{equation}
\mathbf{X}_{4}=\mathbf{H}=\left(  y-Y\right)  \partial_{y}+\left(  z-Z\right)
\partial_{z} \label{TypeIIIFullSpaceCVF1}%
\end{equation}%
\begin{align}
\mathbf{X}_{5}  &  =\left\{  \frac{k}{4}\left[  \left(  y-Y\right)
^{2}-\left(  z-Z\right)  ^{2}\right]  -1\right\}  \partial_{y}+\nonumber\\
& \nonumber\\
&  +\frac{k}{2}\left(  y-Y\right)  \left(  z-Z\right)  \partial_{z}
\label{TypeIIIFullSpaceCVF2}%
\end{align}%
\begin{align}
\mathbf{X}_{6}  &  =\frac{k}{2}\left(  y-Y\right)  \left(  z-Z\right)
\partial_{y}+\nonumber\\
& \nonumber\\
&  +\left\{  \frac{k}{4}\left[  \left(  z-Z\right)  ^{2}-\left(  y-Y\right)
^{2}\right]  -1\right\}  \partial_{z}. \label{TypeIIIFullSpaceCVF3}%
\end{align}
The $\mathbf{X}_{4},\mathbf{X}_{5},\mathbf{X}_{6}$ are \emph{proper} and
\emph{gradient} ICVFs and the conformal factors are given by
\begin{equation}
\phi(\mathbf{X}_{1})=\phi(\mathbf{X}_{2})=\phi(\mathbf{X}_{3})=0
\label{ConfFactorsEuclidean1}%
\end{equation}%
\begin{equation}
\phi(\mathbf{X}_{4})=\left\{  1-\frac{k}{4}\left[  (y-Y)^{2}+(z-Z)^{2}\right]
\right\}  N \label{ConfFactorsEuclidean2}%
\end{equation}%
\begin{equation}
\phi(\mathbf{X}_{5})=kN(y-Y),\qquad\phi(\mathbf{X}_{6})=kN\left(  z-Z\right)
. \label{ConfFactorsEuclidean3}%
\end{equation}
The 2d manifold $\mathbf{x}\wedge\mathbf{u}=\mathbf{0}$ is of \emph{constant
curvature} and the metric (\ref{MetricTimelikeSurfaceConstantCurvature1}) is
\begin{equation}
ds^{2}=dx^{2}-C^{2}dt^{2}+\frac{S^{2}}{V^{2}}\frac{dy^{2}+dz^{2}}{\left\{
1+\frac{\epsilon}{4V^{2}}\left[  \left(  y-Y\right)  ^{2}+\left(  z-Z\right)
^{2}\right]  \right\}  ^{2}} \label{MetricSpacelikeSurfaceConstantCurvature1}%
\end{equation}
where $S(t,x)$ and $Y(t),$ $Z(t),$ $V(t)$ are now arbitrary functions of $t$
and $\epsilon=\pm1$ ($\neq0$) corresponds to the constant curvature of the
hypersurfaces $x,t=$const.

Similarly with type II we define the function $E(t,y,z)$ according to
($k=\epsilon/V^{2}$)
\begin{equation}
E(t,y,z)=V\left\{  1+\frac{k}{4}\left[  \left(  y-Y\right)  ^{2}+\left(
z-Z\right)  ^{2}\right]  \right\}  \label{SpacelikeFunctionE1}%
\end{equation}
with
\begin{equation}
N(t,y,z)=\frac{1}{E(t,y,z)} \label{FunctionN2}%
\end{equation}
and the metric becomes
\begin{equation}
ds^{2}=dx^{2}-C^{2}dt^{2}+\frac{S^{2}}{E^{2}}\left(  dy^{2}+dz^{2}\right)  .
\label{MetricSpacelikeSurfaceConstantCurvature2}%
\end{equation}
For completeness we give the corresponding expressions for the ICVFs and the
metric for the case where the curvature of $\mathbf{x}\wedge\mathbf{u}%
=\mathbf{0}$ vanishes
\begin{equation}
\mathbf{X}_{1}=\mathbf{M}_{yz}=(z-Z)\partial_{y}-\left(  y-Y\right)
\partial_{z} \label{TypeIIIFlatSpaceKVF1}%
\end{equation}%
\begin{align}
\mathbf{X}_{2}  &  =\left[  \left(  y-Y\right)  ^{2}-\left(  z-Z\right)
^{2}\right]  \partial_{y}+\nonumber\\
& \nonumber\\
&  +2\left(  y-Y\right)  \left(  z-Z\right)  \partial_{z}
\label{TypeIIIFlatSpaceKVF2}%
\end{align}%
\begin{align}
\mathbf{X}_{3}  &  =2\left(  y-Y\right)  \left(  z-Z\right)  \partial
_{y}+\nonumber\\
& \nonumber\\
&  +\left[  \left(  z-Z\right)  ^{2}-\left(  y-Y\right)  ^{2}\right]
\partial_{z} \label{TypeIIIFlatSpaceKVF3}%
\end{align}%
\begin{equation}
\mathbf{X}_{4}=\mathbf{H}=\left(  y-Y\right)  \partial_{y}+\left(  z-Z\right)
\partial_{z} \label{TypeIIIFlatSpaceCVF2}%
\end{equation}%
\begin{equation}
\mathbf{X}_{5}=\partial_{y},\quad\mathbf{X}_{6}=\partial_{z}.
\label{TypeIIIFlatSpaceCVF23}%
\end{equation}
The conformal factors are
\begin{equation}
\phi(\mathbf{X}_{1})=\phi(\mathbf{X}_{2})=\phi(\mathbf{X}_{3})=0
\label{ConfFactorsTypeIIIFlat1}%
\end{equation}%
\begin{equation}
\phi(\mathbf{X}_{4})=-1 \label{ConfFactorsTypeIIIFlat2}%
\end{equation}%
\begin{equation}
\phi(\mathbf{X}_{5})=-\frac{2(y-Y)}{\left(  y-Y\right)  ^{2}+\left(
z-Z\right)  ^{2}} \label{ConfFactorsTypeIIIFlat3}%
\end{equation}%
\begin{equation}
\phi(\mathbf{X}_{6})=-\frac{2(z-Z)}{\left(  y-Y\right)  ^{2}+\left(
z-Z\right)  ^{2}} \label{ConfFactorsTypeIIIFlat4}%
\end{equation}
and the metric function $E(t,y,z)$ assumes the form
\begin{equation}
E(t,y,z)=\frac{1}{N(t,y,z)}=\frac{1}{4}\left[  \left(  y-Y\right)
^{2}+\left(  z-Z\right)  ^{2}\right]  . \label{FunctionE3}%
\end{equation}
In contrast with the previous case, the spacetime
(\ref{MetricTimelikeSurfaceConstantCurvature2}) does not allow the existence
of conserved currents and quantities constructed from null vector fields
\textquotedblleft living\textquotedblright\ in $\mathbf{x}\wedge
\mathbf{u}=\mathbf{0}$ due to the positive-definite character of the
quasi-symmetric 2d metric $p_{ab}$. However we can check for a flux-free
solution which implies the \textquotedblleft temporal\textquotedblright%
\ constraints $G_{\ \alpha}^{0}=0$
\begin{equation}
C\left(  ES_{,tx}-S_{,x}E_{,t}\right)  +C_{,x}\left(  SE_{,t}-ES_{,t}\right)
=0 \label{EuclideanConstraint1}%
\end{equation}%
\begin{equation}
CS\left(  EE_{,ty}-E_{,t}E_{,y}\right)  +EC_{,y}\left(  ES_{,t}-SE_{,t}%
\right)  =0. \label{EuclideanConstraint2}%
\end{equation}%
\begin{equation}
CS\left(  EE_{,zt}-E_{,t}E_{,z}\right)  +EC_{,z}\left(  ES_{,t}-SE_{,t}%
\right)  =0 \label{EuclideanConstraint3}%
\end{equation}
and the associated \textquotedblleft spatial\textquotedblright\ constraints
$G_{\ \beta}^{\alpha}=0$
\begin{equation}
SC_{,yx}-C_{,y}S_{,x}=0 \label{EuclideanConstraint4}%
\end{equation}%
\begin{equation}
SC_{,zx}-C_{,z}S_{,x}=0 \label{EuclideanConstraint5}%
\end{equation}%
\begin{equation}
C_{,y}E_{,z}+C_{,z}E_{,y}+EC_{,zy}=0. \label{EuclideanConstraint6}%
\end{equation}
We can verify that the \emph{general solution} of (\ref{EuclideanConstraint1}%
)-(\ref{EuclideanConstraint6}) is
\begin{equation}
C=\frac{S\left[  \ln\left(  S/E\right)  \right]  _{,t}}{\sqrt{\epsilon+F(t)}}
\label{SolutionFuncBSpacelikeSurfConstantCurvat2}%
\end{equation}
with $F(t)$ an arbitrary function and $E(t,y,z)$ is given in
(\ref{SpacelikeFunctionE1}) or (\ref{FunctionE3}). It becomes evident that
also in this type, the existence of the $\mathcal{IC}(\mathbf{X}_{A})$
intrinsic conformal algebra is a direct consequence of the general solution
(\ref{SolutionFuncBSpacelikeSurfConstantCurvat2}), (\ref{SpacelikeFunctionE1})
or (\ref{FunctionE3}).

Using the same arguments, the directional \textquotedblleft
pressures\textquotedblright\ $p_{\alpha}$ are, in general, not equal and the
fluid is anisotropic. This, however, does not exclude \emph{a priori} a
perfect fluid (not tilted) solution once the consistency of the integrability
conditions is established. In addition, the existence of the general solution
of (\ref{EuclideanConstraint1})-(\ref{EuclideanConstraint6}) is
\emph{equivalent} with the fact that the $x-$slices are \emph{conformally
flat} and timelike which indicates a \emph{10-dimensional algebra} of ICVFs
$\mathbf{\Upsilon}$ which will give rise to conserved quantities along null
geodesics of the form $l^{a}=f_{0}u^{a}+f_{1}y^{a}+f_{2}z^{a}$ where
$f_{0},f_{1},f_{2}$ are some functions satisfying the orthonormality condition
$f_{0}^{2}=f_{1}^{2}+f_{2}^{2}$ and the geodesic assumption $\left(
f_{0}u^{a}+f_{1}y^{a}+f_{2}z^{a}\right)  _{;b}l^{b}=0$.

Summarizing the results of Sections II and III regarding the conformal
flatness of the $\mathbf{x}_{\perp}$ distribution in both types II and III, we
can speculate that

\noindent\emph{a spacetime with metric (\ref{GeneralDiagonalMetric1}) is
foliated with conformally flat 3-dimensional hypersurfaces iff a 6-dimensional
subalgebra of ICVFs exists acting on 2d submanifolds and the }$\mathbf{u}%
_{\perp}$ \emph{or} $\mathbf{x}_{\perp}$\emph{ distributions are almost
(1+2)-decomposable (in the spirit of the arguments in
\cite{Apostolopoulos:2016nno}). Equivalently iff a 6-dimensional subalgebra of
ICVFs exists acting on 2d (pseudo)-Riemannian manifolds and the Ricci tensor
shares a common basis of eigenvectors with shear }$\sigma_{ab}$\emph{ and the
electric part }$E_{ab}$ \emph{of the Weyl tensor. }

\section{Conclusions}

It should be noticed that the existence of the 6-dimensional algebra of ICVFs
acting on 2d manifolds is \emph{independent} from the geodesic assumption of
the unit spacelike vector field $x^{a}$ and the form of the metrics
(\ref{MetricTimelikeSurfaceConstantCurvature3}),
(\ref{MetricSpacelikeSurfaceConstantCurvature2}) is altered only by an
arbitrary function in the $g_{xx}-$component with a subsequent change in the
dynamics. As such, the structure of the class of spacetimes presented in this
paper can be regarded as a generalization of the (irrotational) Locally
Rotationally Symmetric (LRS) geometries without any global isometry
containing, however, these models as special cases \cite{vanElst:1995eg}.

An interesting aspect of the analysis of the preceding sections is the
existence of \emph{infinite conserved quantities} along null geodesics
originated from the ICVFs admitted by the 2d submanifold (in type II) or the
$\mathbf{x}_{\perp}-$submanifold (in types II and III). It could be therefore
enlightening the determination of the 10-dimensional algebra of ICVFs due to
the emerged conformal flatness of the hypersurfaces $x=$const. when the fluid
is flux-free $q^{a}=0$ and its anisotropy is described only in terms of the 3
principal inhomogeneous \textquotedblleft pressures\textquotedblright%
\ $p_{\alpha}$ (or equivalently when the Einstein tensor $G_{\ b}^{a}$ is diagonal).

As we have seen, a perfect fluid solution was not excluded a priori. In this
direction it would be interesting to allow the inclusion of a cosmological
constant $\Lambda$ similar to the case of the Szekeres models
\cite{Barrow:1984zz} or the Petrov type I silent universes
\cite{vandenBergh:2004tfa} where exact solutions has been shown to exist.
Furthermore the models of type II
(\ref{MetricTimelikeSurfaceConstantCurvature3}),
(\ref{SolutionFuncBTimelikeSurfConstantCurvat1}) with (\ref{FunctionE1}) and
(\ref{FunctionE2}) or type III (\ref{MetricSpacelikeSurfaceConstantCurvature2}%
), (\ref{SolutionFuncBSpacelikeSurfConstantCurvat2}) with
(\ref{SpacelikeFunctionE1}) and (\ref{FunctionE3}) could be also relevant of
studying the effect of \emph{small} \emph{anisotropic} and
\emph{inhomogeneous} \textquotedblleft pressures\textquotedblright\ to the
expansion dynamics either as the relic of various physical sources
\cite{Barrow:1997sy} or as the result of backreaction terms of the density
fluctuations \cite{Floerchinger:2014jsa, Blas:2015tla} provided that the use
of purely phenomenological laws governing the appearance of \textquotedblleft
pressures\textquotedblright\ is consistent with the kinetic theory approach of
the fluid thermodynamics.

Relaxing the flux-free restrictions (equations (\ref{Constraint1}%
)-(\ref{Constraint3}) or (\ref{EuclideanConstraint1}%
)-(\ref{EuclideanConstraint3})) opens the possibility that exact tilted
perfect fluid solutions could be found for the spacetimes presented in this
paper. Unlike the symmetric Lema\^{\i}tre-Tolman-Bondi (LTB) subclass
\cite{tb} (or the plane/hyperbolic analogues) where a tilted (twisted) perfect
fluid solution cannot exist \cite{vanElst:1995eg} (due to the locally
rotational symmetry), it is far from obvious that the \emph{intrinsic locally
rotational symmetry} induced from the ICVF could be strong enough to forbid a
non-comoving perfect fluid interpretation.

We emphasize that every attempt to assign a dynamical (vacuum or non-vacuum)
interpretation to the spacetimes presented in this paper must take into
account the induced (non-symmetry) integrability conditions. This can be done
by examining whether a suitable set of initial data evolves consistently which
is equivalent to demand that the constraints (spatial divergence and curl
equations encoded in the set of the initial data), are consistent with the
evolution equations hence, they are preserved identically along the timelike
congruence $u^{a}$ \emph{without} imposing new geometrical, kinematical or
dynamical restrictions \cite{Maartens:1996hb, vanElst:1996zs}. Therefore it is
necessary to formulate covariantly the necessary and sufficient conditions,
coming from the existence of the symmetry, and study their consequences in the
dynamics. All the above we believe that are physically sound and require
further investigation.

\end{document}